\documentclass[twocolumn]{article}
\usepackage[T1]{fontenc}
\usepackage[latin9]{inputenc}
\usepackage{amsmath}
\usepackage{amssymb}
\usepackage{graphicx}
\usepackage{CJK}
\usepackage{amsmath}
\usepackage{cases}
\usepackage{indentfirst}
\usepackage{cite}

\newenvironment{sequation}{\begin{equation}\small}{\end{equation}}
\newenvironment{tequation}{\begin{equation}\tiny}{\end{equation}}
\begin{document}
\title{Minimal Dominating Set problem studied by simulated annealing and cavity method:Analytics and population dynamics}
\author{$Yusupjan Habibulla$ \\School of Physics and Technology, xinjiang university, Sheng-Li Road 14,\\  Urumqi 830046, China}

\maketitle
\tableofcontents
\begin{abstract}
The minimal dominating Set (MDS) problem is a prototypical hard combinatorial optimization problem. Two years ago we studied this problem by cavity method. Although we get the solution of a given graph, which gives very good estimation of minimal dominating size, but we don't know whether we get the ground state solution and how many solutions exist in the ground state. For this purpose, last year we continue to develop the one step replica symmetry breaking (RSB) theory to find the ground state energy of the MDS problem. Finally we find that 1) The MDS problem solution space has both condensation transition and cluster transition on regular Random (RR) graph and we prove this by simulated annealing dynamical process. 2) We developed zero temperature Survey Propagation (SP) algorithm on ER graph to estimate the ground state energy and to get Survey Propagation Decimation (SPD) algorithm with good results same as BPD algorithm.
\end{abstract}
\section{ Introduction}
The statistical physics of spin glass systems has widely application in optimization problem (for example the minimal vertex cover problem, minimum feedback vertex set and minimal dominating set,etc), and  satisfiability problem ( for example k-sat,XOR-sat). Many combinatorial optimization problem in computer science could be mapped into appropriate random spin glass model. It mainly use the cavity method to estimate the occupy probability of each nodes, and upon this probability to construct one solution of the given problem. If the graph no contains any shortest distance cycles or has tree-like local structure, we can find the stable point of the iterative cavity equations, namely we can study such systems by cavity method.\\
At present, our research on the spin glass system mainly concentrated on replica symmetric and one step replica symmetric breaking level. Each level research corresponding to different purpose of the combinatorial optimization problem, every combinatorial optimization problem almost answer following some question:(1)can we find the smallest set of the given graph to satisfy the problem?(2)how much configuration (or solution) of the given graph satisfy the problem?(3)for the given real number $x_{c} (0<x_{c}<1)$, can we judge if there exist a solution of size not exceeding $x_{c}N$ (in there, $N$ represent the node number of the given graph)?\\
Previously we have tried to answer the question one of minimal dominating set problem through cavity method and get good results\cite{1,2}. Now the following work inspired by \cite{3}, in this work we mainly answer the last question of minimal dominating set problem through cavity method too. The following bounds on minimal dominating sets are known(Haynes, Hedetniemi and Slater 1998a, Chapter 2)\cite{4}, one vertex can dominate at most $\Delta$ other vertices; therefore $\gamma(G) \geq n/(1 + \Delta)$, in there the $\gamma(G)$ represents the size of the minimal dominating set. So $x_{c}=1/(1+\Delta)$, for large regular random graph we find that a sharp threshold value $x_{c}\geq 1/(1+\Delta)$ exists, this value closely related to computational complexity. When $x>x_{c}$, we can construct a minimal dominating set of size $\leq xN$ for a given graph; while when $x<x_{c}$, we can not construct any minimal dominating set of size $\leq xN$. When $x>x_{c}$, simulated annealing algorithm relatively easy to give a solution of such a minimal dominating set; however, when x close to $x_{c}$, the search complexity increases dramatically, and trend to infinity when $x=x_{c}$, so in finite computation time it can't give a solution  of such a minimal dominating set. At zero temperature, we use survey propagation to find the threshold value $x_{c}$ for ER random graph too.\\
A vertex minimal dominating set(MDS)\cite{4} of a given graph G is a set of vertices D such that every vertex of G is either in D or at least one neighbor in D.
MDS is a very important branch of graph theory, it is widely studied by mathematician and physician in theoretical and algorithm point of view. It is also a very important to computer science and artificial intelligence, it has widely application in complex network system \cite{5,6,7,8,9,10,11,12}.\\
The MDS problem was studied from the 1950s onwards, But the more and more researchers work on the MDS after mid-1970s, There are more than 400 papers related to MDS problem, in which almost papers related to following three factors:(1)the diversity of applications to both real-world and other mathematical
"covering" or "location" problems;(2)the wide variety of domination parameters that can be defined;(3)the
NP-completeness of the basic domination problem, its close and "natural" relationships to other NP-complete problems, and the subsequent interest in finding polynomial time solutions to domination problems in special classes of graphs\cite{13}. MDS problem is a nondeterministic polynomial-complete(NP-complete) optimization problem\cite{4}, So finding exact solution is extremely difficult task in general.Even we hard to find the approximate MDS solution of a given graph. There are some heuristic algorithms\cite{5,6,7,9,14,15} and statistical physics algorithm\cite{1,2} to solve the MDS problem, but the only small part of this work related to solution space structure, bounds (threshold value $x_{c}$) and  size of MDS problem.\\
In this work we use one step replica symmetry breaking theory of statistical physics to study the solution space of minimal dominating set problem.
we organize the paper as fellows: First, in section 2 we recall the replica symmetry theory of spin glass in order to convenient to understanding the replica symmetry breaking theory.we present the BP equation and thermodynamic quantities.  The first main part of our work is put in the section 3 we introduce the one step replica symmetry breaking theory and thermodynamic quantities.In the second part of this section we derive the one step replica symmetry breaking theory and thermodynamic quantities at $y=\beta$, and introduce simulated annealing dynamical process for the MDS, in the third part we introduce the population dynamics for this case in detail. The second main part of our work is put in the section 4. In this Section we derive the warning propagation,Survey propagation and Survey propagation Decimation.Finally in Section 5 we conclusion our results.

\section{ Replica Symmetry}
In order to estimate the MDS for a given graph by the way of mean field theory, we must have the partition function for the given problem. we now introduce a partition function Z as
\begin{equation}
Z=\sum_{\underline{c}}\prod_{i\in W}{e^{-\beta c_{i}}}[1-(1-c_{i})\prod_{k\in\partial i}(1- c_{k})]
\end{equation}
in there $\underline{ c}\equiv( c_{1}, c_{2},......, c_{n})$denotes one of the $2^{n}$ possible configurations, $ c_{i}=+1$ if node $i$ be occupied and $ c_{i}=0$ if otherwise, $\beta$ is inverse temperature,and $\partial i$ denotes the neighbor nodes of node $i$,The partition function therefore only takes into account all the dominating sets.\\
we use RS mean field theory such as the Bethe-Peierls approximation\cite{16} or partition function expansion\cite{17,18} to solve the above spin glass model. we set cavity message $p_{i\rightarrow j}^{(c_{i},c_{j})}$ on the every edge,and these messages must satisfy following equation

\begin{equation}
p_{i\rightarrow j}^{(c_{i},c_{j})}=\frac{e^{-\beta c_{i}}\prod\limits_{k\in\partial i\backslash j}\sum\limits_{c_{k}}p_{k\rightarrow i}^{( c_{k}, c_{i})}-\delta_{ c_{i}}^{0}\delta_{ c_{j}}^{0}\prod\limits_{k\in\partial i\backslash j}p_{k\rightarrow i}^{(0,0)}}{\sum\limits_{\acute{ c}_{i},\acute{ c}_{j}}e^{-\beta \acute c_{i}}\prod\limits_{k\in\partial i\backslash j}\sum\limits_{\acute c_{k}}p_{k\rightarrow i}^{( \acute c_{k}, \acute c_{i})}-\prod\limits_{k\in\partial i\backslash j}p_{k\rightarrow i}^{(0,0)}}
\end{equation}
this equation called Belief-Propagation (BP) equation.In there the Kronecker symbol $\delta_{m}^{n}=1$ if $m=n$ and $\delta_{m}^{n}=0$ if otherwise. The cavity message $p_{i\rightarrow j}^{(c_{i},c_{j})}$ represents the joint probability that node $i$ is in occupation state $c_{i}$ and its adjacent node $j$ is in occupation state $c_{j}$ when the constraint of node $j$ is not considered.The marginal probability $p_{i}^{c}$ of node $i$ is expressed as
\begin{equation}
p_{i}^{c}=\frac{e^{-\beta c}\prod\limits_{j\in\partial i}\sum\limits_{c_{j}}p_{j\rightarrow i}^{( c_{j}, c)}-\delta_{ c}^{0}\prod\limits_{j\in\partial i}p_{j\rightarrow i}^{(0,0)}}{\sum\limits_{c_{i}}e^{-\beta c_{i}}\prod\limits_{j\in\partial i}\sum\limits_{c_{j}}p_{j\rightarrow i}^{(c_{j}, c_{i})}-\prod\limits_{j\in\partial i}p_{j\rightarrow i}^{(0,0)}}
\end{equation}

finally the free energy can be calculated by mean field theory
\begin{equation}
F_{0}=\sum_{i=1}^{N}F_{i}-\sum_{(i,j)=1}^{M}F_{(i,j)}
\end{equation}
in there

\begin{equation}
F_{i}=-\frac{1}{\beta}\ln[\sum\limits_{c_{i}}e^{-\beta c_{i}}\prod\limits_{j\in\partial i}\sum\limits_{c_{j}}p_{j\rightarrow i}^{(c_{j}, c_{i})}-\prod\limits_{j\in\partial i}p_{j\rightarrow i}^{(0,0)}]
\end{equation}

\begin{equation}
F_{(i,j)}=-\frac{1}{\beta}\ln[\sum_{ c_{i}, c_{j}}p_{i\rightarrow j}^{( c_{i}, c_{j})}p_{j\rightarrow i}^{( c_{j}, c_{i})}]
\end{equation}
In there the $F_{i}$ denotes the free energy of function node $i$, the $F_{(i,j)}$ denotes the free energy of the edge $(i,j)$.We iterate the BP equation until converge to one stable point to calculate the mean free energy $f\equiv F/N$ and the energy density $\omega=1/N\sum_{i}p_{i}^{+1}$ by equation (3) and (4). The entropy density calculates as $s=\beta(\omega-f)$.
\section{One Step Replica Symmetry Breaking theory}
In this section we will introduce one step replica symmetry breaking theory of the spin glass by graph expansion method. Firstly we introduce the generalized partition function, grand free energy and survey propagation for general case, in order to rising the simulation speed
we must simplify these equations,so secondly we derive the simplified equations at $y=\beta$ for minimal dominating set problem,
and then introduce the numerical simulation process of population dynamics.
\subsection{ General One step Replica Symmetry Breaking Theory}
At higher temperature,the thermodynamic microscopic state that consist of some higher energy configuration decide the statistical physics property of the given system, and the subspace of this microscopic state is ergodic. But in the lower temperature the microscopic state is not ergodic anymore, it is divided into several subspaces and the contribution of this
subspaces to the equilibrium property is not same. If we select the energy function as order parameter, We don't know how is this subspace like
 and how do it evolution. So scientist select the free energy function as order parameter to develop the one step replica symmetry breaking theory.
we define the generalized partition function $\Xi$ as
\begin{equation}
\Xi(y;\beta)=\sum_{\alpha}e^{-yF_{0}^{\alpha}(\beta)},
\end{equation}
While $\alpha$ is denoted as the macroscopic states where the free energies achieves the minimum value, $F_{0}^{\alpha}$ has the form as below.
For the sake of simplification we use $i\rightarrow j$ to represent the $i\rightarrow (i,j)$,

\begin{equation}
F_{0}^{(\alpha)}=\sum_{i}f_{i} - \sum_{(i,j)}f_{(i,j)}.
\end{equation}
Four cavity messages $p_{i\rightarrow j}\equiv (p_{i\rightarrow j}^{0,0},p_{i\rightarrow j}^{0,1},p_{i\rightarrow j}^{1,0},p_{i\rightarrow j}^{1,1},)$ are defined on any edges $(i,j)$ in a given graph instance, and the cavity messages $p_{i\rightarrow j}$ averaged on solution clusters is denoted as $P_{i\rightarrow j}(p_{i\rightarrow j})$, which have the iteration equation as
\begin{tequation}
P_{i\rightarrow j}(p)=\frac{\prod_{k\in\partial i\backslash j}\int\mathcal{D}p_{k\rightarrow i}P_{k\rightarrow i}(p)e^{-yf_{i\rightarrow j}}\delta(p_{i\rightarrow j}-I_{i\rightarrow j}[p_{\partial i\backslash j}])}{\prod_{k\in\partial i\backslash j}\int\mathcal{D}p_{k\rightarrow i}P_{k\rightarrow i}(p)e^{-yf_{i\rightarrow j}}}
\end{tequation}
While $I_{i\rightarrow j}[p_{\partial i\backslash j}]$ is a short-hand notation of the message updating equations of Eqs.2.
The generalized free energy density $g_{0}$ as
\begin{equation}
g_{0}\equiv \frac{G_{0}}{N}=\frac{\sum_{i}g_{i}-\sum_{(i,j)}g_{(i,j)}}{N}
\end{equation}
where
\begin{equation}
g_{i}=\frac{1}{y}\ln[\prod_{j\in\partial i}\int\mathcal{D}p_{i\rightarrow j}P_{i\rightarrow j}(p)e^{-yf_{i}}]
\end{equation}
\begin{sequation}
g_{(i,j)}=\frac{1}{y}\ln[\int\int\mathcal{D}p_{i\rightarrow j}\mathcal{D}p_{j\rightarrow i}P_{i\rightarrow j}(p)P_{j\rightarrow i}(p)e^{-yf_{(i,j)}}]
\end{sequation}

We further have the mean free energy density $<f>$ as

\begin{equation}
<f>\equiv \frac{F}{N}=\frac{\sum_{i}<f_{i}>-\sum_{(i,j)}<f_{(i,j)}>}{N}
\end{equation}
in there
\begin{equation}
<f_{i}>=\frac{\prod_{j\in\partial i}\int\mathcal{D}p_{i\rightarrow j}P_{i\rightarrow j}(p)e^{-yf_{i}}f_{i}}{\prod_{j\in\partial i}\int\mathcal{D}p_{i\rightarrow j}P_{i\rightarrow j}(p)e^{-yf_{i}}}
\end{equation}
\begin{tequation}
<f_{(i,j)}>=\frac{\int\int\mathcal{D}p_{i\rightarrow j}\mathcal{D}p_{j\rightarrow i}P_{i\rightarrow j}(p)P_{j\rightarrow i}(p)e^{-yf_{(i,j)}}f_{(i,j)}}{\int\int\mathcal{D}p_{i\rightarrow j}\mathcal{D}p_{j\rightarrow i}P_{i\rightarrow j}(p)P_{j\rightarrow i}(p)e^{-yf_{(i,j)}}}
\end{tequation}
Finally we have the complexity with mean free energy $<f>$ and generalized free energy $g$, we derive the complexity as
$\sum(y)=y(<f>-g)$.
\subsection{Dynamical and Condensation Transitions at $y=\beta$}
We consider the case of $y=\beta$ to determine the clustering transition and condensation transition. For the sake of simplify the formula of Replica
Symmetry Breaking and raising the simulation speed we set two kind of messages.
We define the average message $\bar{p}_{i\rightarrow j}^{( c_{i}, c_{j})}$ and the conditional message $P_{i\rightarrow j}^{( c_{i}, c_{j})}(p_{i\rightarrow j}^{( c_{i}, c_{j})}|\bar{p}_{i\rightarrow j}^{( c_{i}, c_{j})})$ as.
\begin{equation}
\bar{p}_{i\rightarrow j}^{( c_{i}, c_{j})}=\int\mathcal{D} p P_{i\rightarrow j}[p] p_{i\rightarrow j}^{( c_{i}, c_{j})}
\end{equation}
\begin{equation}
P_{i\rightarrow j}^{( c_{i}, c_{j})}(p_{i\rightarrow j}^{( c_{i}, c_{j})}|\bar{p}_{i\rightarrow j}^{( c_{i}, c_{j})})=\frac{p_{i\rightarrow j}^{( c_{i}, c_{j})}P_{i\rightarrow j}[p_{i\rightarrow j}]}{\bar{p}_{i\rightarrow j}^{( c_{i}, c_{j})}}
\end{equation}
where the variable node $(i,j)$ only participate the interaction $i$,the conditional message
$P_{i\rightarrow j}^{( c_{i}, c_{j})}(p_{i\rightarrow j}^{( c_{i}, c_{j})}|\bar{p}_{i\rightarrow j}^{( c_{i}, c_{j})})$
represents the macroscopic conditional probability of the cavity probabilistic distribution function $p_{i\rightarrow j}^{( c_{i}, c_{j})}$ when we investigate the pair node $(i,j)$ in the state$( c_{i}, c_{j})$. From these two equations and BP equations we can derive the new update rules for these two messages(average message and conditional message) as
\begin{equation}
\bar{p}_{i\rightarrow j}^{(c_{i},c_{j})}=\frac{e^{-\beta c_{i}}\prod\limits_{k\in\partial i\backslash j}\sum\limits_{c_{k}}\bar{p}_{k\rightarrow i}^{( c_{k}, c_{i})}-\delta_{ c_{i}}^{0}\delta_{ c_{j}}^{0}\prod\limits_{k\in\partial i\backslash j}\bar{p}_{k\rightarrow i}^{(0,0)}}{\sum\limits_{\acute{ c}_{i},\acute{ c}_{j}}e^{-\beta \acute c_{i}}\prod\limits_{k\in\partial i\backslash j}\sum\limits_{\acute c_{k}}\bar{p}_{k\rightarrow i}^{( \acute c_{k}, \acute c_{i})}-\prod\limits_{k\in\partial i\backslash j}\bar{p}_{k\rightarrow i}^{(0,0)}}
\end{equation}
\begin{tequation}
\begin{split}
P_{i\rightarrow j}^{( c_{i}, c_{j})}(p_{i\rightarrow j}^{( c_{i}, c_{j})}|\bar{p}_{i\rightarrow j}^{( c_{i}, c_{j})})\\
=\sum_{\underline{ c}_{\partial i\backslash j}}\omega_{i\rightarrow j}^{( c_{i}, c_{j})}&\prod_{k\in i\backslash j}\int\mathcal{D} p_{k\rightarrow i}P_{k\rightarrow i}^{( c_{k}, c_{i})}(p_{k\rightarrow i}^{( c_{k}, c_{i})}|\bar{p}_{k\rightarrow i}^{( c_{k}, c_{i})})\\
\times\delta(p_{i\rightarrow j}^{ c_{i}, c_{j}}-&A_{i\rightarrow j}(p_{\partial i\backslash j}))
\end{split}
\end{tequation}
in there
\begin{tequation}
\omega_{i\rightarrow j}^{(c_{i},c_{j})}=\frac{e^{-\beta c_{i}}\prod_{k\in\partial i\backslash j}\sum_{ c_{k}}\bar{p}_{k\rightarrow i}^{( c_{k}, c_{i})}-\delta_{c_{i}}^{0}\delta_{c_{j}}^{0}\prod_{k\in\partial i\backslash j}\bar{p}_{k\rightarrow i}^{(0,0)}}{\sum_{\underline{ c}_{\partial i\backslash j}}e^{-\beta c_{i}}\prod_{k\in\partial i\backslash j}\sum_{ c_{k}}\bar{p}_{k\rightarrow i}^{( c_{k}, c_{i})}-\prod_{k\in\partial i\backslash j}\bar{p}_{k\rightarrow i}^{(0,0)}}
\end{tequation}
In there $\omega_{i\rightarrow j}^{(c_{i},c_{j})}$ denotes the sampling joint probability,we can calculate the joint probability by average messages,and then using this probability to get one configuration sample of
$\partial i\backslash j\equiv \{ c_{k}:k\in\partial i\backslash j\}$.After this we use this configuration sample to update conditional messages.
Even these survey propagation looks like complicated and hard to program than original survey propagation, but these equations easier to converge than original one.In the same way we can derive the thermodynamic quantities, the total free energy can be expressed as
\begin{equation}
F=\sum_{i}<f_{i}>-\sum_{(i,j)}<f_{(i,j)}>
\end{equation}
in there the $<f_{i}>$ denote the mean free energy of node $i$,it can be expressed as
\begin{sequation}
\begin{split}
<f_{i}>&=-\frac{1}{\beta}\sum_{\underline{ c}_{\partial i}}\omega_{i}^{c}\prod_{j\in i}\int D p_{j\rightarrow i}P_{j\rightarrow i}^{( c_{j}, c_{i})}(p_{j\rightarrow i}^{( c_{j}, c_{i})}|\bar{p}_{j\rightarrow i}^{( c_{j}, c_{i})})\\
&\times\ln\{\sum_{ c_{i}}e^{-\beta c_{i}}\prod_{j\in\partial i}\sum_{ c_{j}}p_{j\rightarrow i}^{( c_{j}, c_{i})}-\prod_{j\in\partial i}p_{j\rightarrow i}^{(0,0)}\}\\
\end{split}
\end{sequation}
In there $\omega_{i}^{c}$ is the probability of sampling, it can be expressed as
\begin{equation}
\omega_{i}^{c}=\frac{e^{-\beta c}\prod_{j\in\partial i}\bar{p}_{j\rightarrow i}^{( c_{j}, c)}-\delta_{c}^{0}\prod_{j\in\partial i}\bar{p}_{j\rightarrow i}^{(0,0)}}{\sum_{ c_{i}}e^{-\beta c_{i}}\prod_{j\in\partial i}\sum_{ c_{j}}\bar{p}_{j\rightarrow i}^{( c_{j}, c_{i})}-\prod_{j\in\partial i}\bar{p}_{j\rightarrow i}^{(0,0)}}
\end{equation}

How to sample the conditional message is very important to get good results, it is core problem in our numerical simulation.
According to the above equation and Appendix A, we determine the state of the selected variables one by one to get one configuration. We select the conditional message by this selected configuration to calculate the mean free energy of node $i$.In the same way we can calculate the free energy of edge $(i,j)$ as

\begin{tequation}
\begin{split}
<F_{(i,j)}>=&-\frac{1}{\beta}\sum_{ c_{i}, c_{j}}\omega_{i\rightarrow j}^{(c_{i}, c_{j})}\int D p_{i\rightarrow j}P_{i\rightarrow j}^{( c_{i}, c_{j})}(p_{i\rightarrow j}^{( c_{i}, c_{j})}|\bar{p}_{i\rightarrow j}^{( c_{i}, c_{j})})\\
&\times\ln\{\sum_{ c_{i}, c_{j}}p_{i\rightarrow j}^{( c_{i}, c_{j})}p_{j\rightarrow i}^{( c_{j}, c_{i})}\}
\end{split}
\end{tequation}
in there
\begin{equation}
\omega_{i\rightarrow j}^{(c_{i}, c_{j})}=\frac{\bar{p}_{i\rightarrow j}^{( c_{i}, c_{j})}\bar{p}_{j\rightarrow i}^{( c_{j}, c_{i})}}{\sum_{ c_{i}, c_{j}}\bar{p}_{i\rightarrow j}^{( c_{i}, c_{j})}\bar{p}_{j\rightarrow i}^{( c_{j}, c_{i})}}
\end{equation}
where we use this above equation to get one pair state $( c_{i}, c_{j})$ configuration of edge $(i,j)$,and then select the conditional messages by this pair state to calculate the free energy of edge $(i,j)$.  The grand free energy of edge $(i,j)$ and node $i$ as
\begin{equation}
g_{(i,j)}=-\frac{1}{\beta}\ln\{\sum_{ c_{i}, c_{j}}\bar{p}_{i\rightarrow j}^{( c_{i}, c_{j})}\bar{p}_{j\rightarrow i}^{( c_{j}, c_{i})}\}
\end{equation}

\begin{equation}
g_{i}=-\frac{1}{\beta}\ln[\sum_{ c_{i}}e^{-\beta c_{i}}\prod_{j\in\partial i}\sum_{ c_{j}}\bar{p}_{j\rightarrow i}^{( c_{j}, c_{i})}-\prod_{j\in\partial i}\bar{p}_{j\rightarrow i}^{(0,0)}]
\end{equation}
we can see that the ground free energy is calculated by average messages very easily. And then the total mean free energy, total mean grand free energy and complexity are calculated as following
\begin{equation}
g_{0}=\frac{G_{0}}{N}=\frac{\sum_{i=1}^{N}g_{i}}{N}-\frac{M}{N}\frac{\sum_{(i,j)=1}^{M}g_{(i,j)}}{M}
\end{equation}
\begin{equation}
f_{0}=\frac{F_{0}}{N}=\frac{\sum_{i=1}^{N}<f_{i}>}{N}-\frac{M}{N}\frac{\sum_{(i,j)=1}^{M}<f_{(i,j)}>}{M}
\end{equation}
\begin{equation}
\sum=\frac{y}{N}(<F_{0}>-G_{0})
\end{equation}
in there $f_{0}$ denotes the mean free energy density and $g_{0}$ denotes the grand free energy density.\\
\begin{figure}[htbp]
  \centering
  \includegraphics[width=8cm,height=3cm]{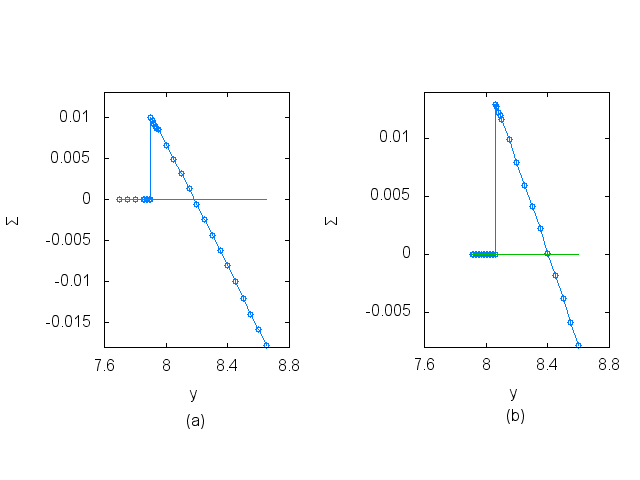}
  \caption{x-axis denotes the inverse temperature,Y-axis denotes the complexity of Regular Random graph that variable degree $C=5$ in left panel (a) and $C=6$ in right panel (b).When $C=5$,the complexity has condensation transition at $\beta=7.90$ and has cluster transition around at $\beta=8.19.$}
\end{figure}
\begin{table}[!hbp]
\caption{the cluster transition of inverse temperature $\beta_{d}$ and condensation transition of inverse temperature $\beta_{c}$ of Regular Random graph}
\begin{tabular}{p{0.45cm}p{0.6cm}p{0.45cm}p{0.45cm}p{0.45cm}p{0.45cm}p{0.45cm}p{0.45cm}p{0.45cm}}
\hline
C & 3 & 4 & 5 & 6&7&8&9&10 \\
\hline
$\beta_{d}$ & 8.06 & 7.81 & 7.9 & 8.06&8.24&8.43&8.63&8.76 \\
\hline
$\beta_{c}\approx$ & 8.25 & 8.04 & 8.19 & 8.41&8.65&8.88&9.11&9.33 \\
\hline
\end{tabular}
\end{table}
Now we discuss the influence of cluster transition to the dynamical property of the regular random graph which mean degree $C=5$ ,the figure.1(a) indicate the relationship between complexity and inverse temperature $\beta$ when $y=\beta$.The complexity jumping to positive when $\beta\approx 7.90$, and then decreasing with the increasing of $\beta$, finally it comes to zero when $\beta\approx 8.19$.We find that the figure.1 basically like with the three body interaction spin glass system.\\
\begin{figure}[h]
  \centering
  \includegraphics[width=8cm,height=5cm]{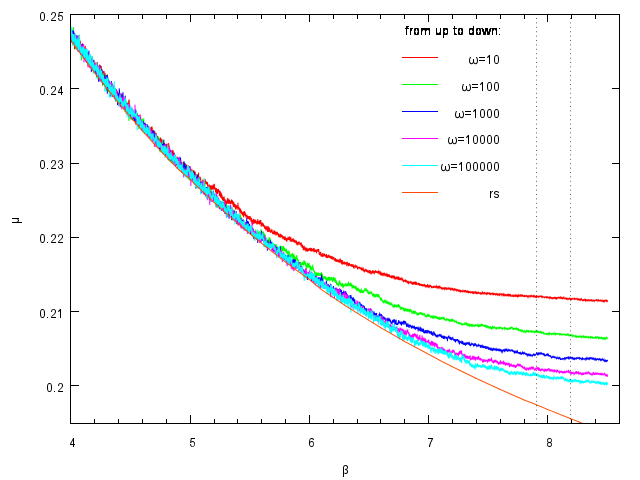}
  \caption{x-axis denotes the inverse temperature, Y-axis denotes the energy density of the Regular Random graph, it's mean connectivity C=5 and includes 1000 variables. The $\omega$ denotes the iterations in every step. }
\end{figure}
The figure.2 indicate the evolution of energy density with the change of inverse temperature $\beta$ in the simulation annealing process. In simulation annealing process,firstly we carry out single spin heat bath dynamical process enough steps at inverse temperature $\beta=1$, so then we can get one equilibrium microscopic configuration sample. Start from this equilibrium microscopic configuration, increase the inverse temperature with the certain speed, every time we increase the inverse temperature $\beta$ with $\delta\beta=0.001$.In every new inverse temperature $\beta$, we firstly carry out $\omega$ steps ( in every step we try to flip every spin of the system on average), and then record the energy density of the microscopic configuration.We totally simulate 96 independent path for the given MDS configuration, so then we get the average value of the energy density of these 96 paths at every inverse temperature.\\
The figure.2 show that, if the speed of the increasing of inverse temperature is very fast, then the energy density deviate from the value of the energy density of the mean field theory predicted at lower inverse temperature.If we slow down the (or increasing the waiting time) speed of the inverse temperature, so then the energy density deviate from the value of the energy density that predicted by mean field theory when the inverse temperature more high.The reason of leading this difference between simulation and theoretical results is, the characteristic relaxation time become more and more longer with the increasing of inverse temperature, if the characteristic relaxation time exceeds the average simulated annealing waiting time, the microscopic configuration sample that getting by simulated annealing simulation process is not equilibrium configuration of the system, and the energy of these configuration higher than the average energy of the equilibrium configuration.\\
Furthermore, even the waiting time $\omega\rightarrow\infty$ at every $\beta$, the average energy density of the simulated annealing impossible to same with the equilibrium energy density when $\beta > \beta_{d}$.  We can correctly compute the equilibrium mean energy density by belief propagation equation when $\beta_{d}<\beta<\beta_{c}$. From the equation.2 we obviously see the difference of simulated annealing mean energy density and equilibrium mean energy density in the range $\beta_{d}<\beta<\beta_{c}$.\\
The intrinsically reason of this difference is that the microscopic equilibrium configuration space appear ergodicity-breaking.In the simulated annealing simulation, when the inverse temperature $\beta$ up to $\beta_{d}$, the microscopic configuration of this system be captured by thermodynamical macroscopic state $\alpha$, even the waiting time $\omega$ very long, the system can not escape from this thermodynamical macroscopic state. So when the inverse temperature further increasing, the system still stay in the microscopic configuration subspace of the thermodynamical macroscopic state $\alpha$, so the simulated annealing mean energy density is the energy density of thermodynamical macroscopic state $\alpha$ at the inverse temperature $\beta$. This mean energy density higher than the equilibrium energy density, the reason of this, when inverse temperature $\beta$ exceed the $\beta_{d}$, the another thermodynamical macroscopic states which has further small energy density determine the equilibrium statistical property of this system.\\

\begin{figure}[h]
  \centering
  \includegraphics[width=8cm,height=5cm]{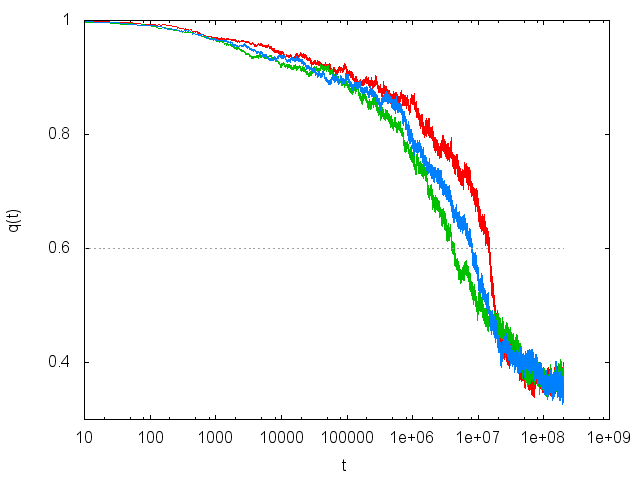}
  \caption{We implement single spin heat bath dynamical process on Regular Random network of the Minimal Dominating Set problem at inverse temperature $\beta=7.5$. This network including $N=1000$ variable nodes. x-axis denotes the evolution log time,Y-axis denotes the overlap. The graph indicates that the overlap between two microscopic configurations of two independent heat bath dynamical process evolves along with time. These three curves set out respectively from different three original microscopic configuration. The horizontal line indicating the overlap value equal to 0.6. }
\end{figure}

When the inverse temperature $\beta$ close to $\beta_{d}$, because of the ergodicity of the microscopic configuration  begin to broken, the dynamical relaxation time will be getting more and more long, and it diverges at $\beta=\beta_{d}$, so the system begin to stay in the non equilibrium state.In order to quantitative compute the characteristic relaxation time, start from the randomly obtained one equilibrium microscopic configuration, perform two completely independent single spin heat bath dynamical process, and record the overlap $q_{t}$ between the two microscopic configurations of the two process at the same time $t$. The figure.3 indicate the three evolution curve of overlap that is getting by starting from three different equilibrium microscopic configuration along with time $t$. These curves are getting on the regular random graph which includes $N=1000$ variable nodes and the variable degree $C=5$, the inverse temperature $\beta=7.5$, it is lower than cluster transition inverse temperature $\beta_{d}=7.9$.\\
The overlap $q_{0}=1$ when $t=0$. when the evolution time enough long, the system totally forget the original state, so that the overlap $q_{t}\rightarrow q^*$. We record the time that the overlap first time decline to $q^{*}=0.6$ as the characteristic relaxation time $\tau$ of one path. We can get the distribution of characteristic relaxation time through simulating many evolution paths, and from this distribution we can obtain the average value and median value of the characteristic relaxation time.The figure.4 indicates the changes of average characteristic relaxation time and median characteristic relaxation time along with inverse temperature $\beta$. For this $C=5$ case, we find that the characteristic relaxation time of this system diverge when $\beta=7.9$.\\
\begin{figure}[htbp]
  \centering
  \includegraphics[width=8cm,height=5cm]{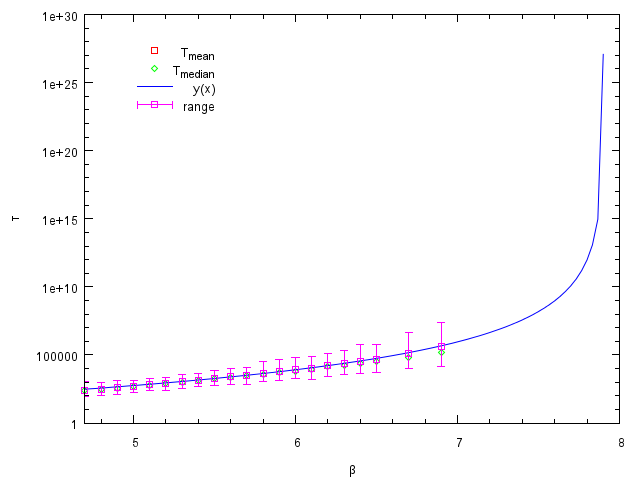}
  \caption{This regular random network include N=1000 nodes,we get 1000 samples of the relaxation time by heat bath dynamical simulation process, and then we calculate the average $\tau_{mean}$ and median $\tau_{median}$ relaxation time of these 1000 relaxation time samples. The solid line is the fit curves of the average relaxation time $\tau_{mean}=\frac{a}{{(\beta^{*}-\beta)}^z}$, in there, $a=(4.75668\pm 0.005399)\times 10^5,\beta ^*=7.9004\pm 0.446,z=6.3125\pm 1.273$. }
\end{figure}
In simulation process, we find that the overlap curve reach a plateau very quickly at every inverse temperature, and then the overlap value vibration in the very small area, but the overlap value of the plateau increase along with the inverse temperature, for example the overlap $q\approx 0.1$  when inverse temperature $\beta=2.0$ and $q\approx 0.35$ when inverse temperature $\beta=7.8$. And when the inverse temperature $\beta$ approach the $\beta_{d}$, the relaxation time is not a gaussian distribution, it has long tail, this leads the average relaxation time not more a right physical quantity. The reason of leading this phenomenon is that, when the inverse temperature $\beta$ approach the $\beta_{d}$, although the equilibrium microscopic configuration also is ergodic, but the microscopic configuration already forms a lot of community, different microscopic configuration community has different characteristic relaxation time. Appearing of different microscopic configuration community can explains that the overlap $q_{t}$ don't decrease as exponential function, but appear a plateau firstly, and then decline very fast. In the logarithmic coordinate, the overlap $q_{t}$ plateau correspond to same community of the microscopic configuration space that explored by two different single spin heat bath dynamical process, and the slump of the overlap $q_{t}$ correspond to that the at least one heat bath dynamical process escaping out from this microscopic configuration community.\\
When the difference of the relaxation time of different evolution path is very big, the Figure.4 indicates that the median of the relaxation time is still a good statistical quantity of the relaxation time.\\
\subsection{Population Dynamics for the one Step Replica Symmetry Breaking at $y=\beta$}
Here we explain the numerical procedure of deriving the thermodynamic quantities such as mean free energy,grand free energy and
 complexity$\sum(y)$ at a given inverse temperature $\beta$ in detail.\\

$\textbf{(1)\hspace{0.5mm} Initialization:}\hspace{1.5mm}$ We construct an array with row size $\mathcal{N}_{0}$ as $\mathcal{S}_{0}$,$\mathcal{S}_{1}$,$\mathcal{S}_{2}$,
$\cdots\mathcal{S}_{\mathcal{N}_{0}-1}$,each of which have two elements of $(\bar{p}_{i\rightarrow j},P_{i\rightarrow j}(p|\bar{p}))$. Each average message $\bar{p}_{i\rightarrow j}$ contains four messages $(\bar{p}_{i\rightarrow j}^{(0,0)},\bar{p}_{i\rightarrow j}^{(0,1)},\bar{p}_{i\rightarrow j}^{(1,0)},\bar{p}_{i\rightarrow j}^{(1,1)})$. But these four messages not independent each other, so we set only two messages $(\bar{p}_{i\rightarrow j}^{(0,0)},\bar{p}_{i\rightarrow j}^{(0,1)})$ to describe the average message.For the conditional message we set different four type eight messages $\{\textcircled{1}p_{i\rightarrow j}(0,0|0,0),p_{i\rightarrow j}(0,1|0,0);\hspace{3mm}\textcircled{2} p_{i\rightarrow j}(0,0|0,1),\\p_{i\rightarrow j}(0,1|0,1);
\hspace{3mm}\textcircled{3} p_{i\rightarrow j}(0,0|1,0),p_{i\rightarrow j}(0,1|1,0);\\
\hspace{3mm}\textcircled{4} p_{i\rightarrow j}(0,0|1,1),p_{i\rightarrow j}(0,1|1,1)\}$. In there $p_{i\rightarrow j}(0,0|0,0)$ represent the probability of node $i$ in state $0$ and node $j$ in state $0$ when we observe the node $i$ in state $0$ and node $j$ in state $0$. Initialize the messages is very important to get correct results, we initialize the messages as
$\{\textcircled{1}p_{i\rightarrow j}(0,0|0,0)=0.5,\hspace{3mm}p_{i\rightarrow j}(0,1|0,0)=0.5;\hspace{6mm}\textcircled{2} p_{i\rightarrow j}(0,0|0,1)=0.0,\hspace{3mm}p_{i\rightarrow j}(0,1|0,1)=1.0;\hspace{6mm}\textcircled{3} p_{i\rightarrow j}(0,0|1,0)=0.0,\hspace{3mm}p_{i\rightarrow j}(0,1|1,0)=0.0;\hspace{3mm}\textcircled{4} p_{i\rightarrow j}(0,0|1,1)=0.0,\hspace{3mm}p_{i\rightarrow j}(0,1|1,1)=0.0\}$.\\

$\textbf{(2)\hspace{0.5mm} Updating:}\hspace{1.5mm}$ There are a lot of sampling scheme\cite{19,20,21}.
In this paper we update the messages with the sampling scheme of Appendix A.\\
$\hspace{1cm}\textbf{(2.1):}\hspace{0.5mm}$ In the single step of iteration, we randomly choose $k-1$ messages from the population.Calculate the new average message by equation(18).\\
$\hspace{1cm}\textbf{(2.2):}\hspace{0.5mm}$After the updating of average message we can get four different samples of conditional messages for the four conditional case by these equations of the Appendix A.\\
$\hspace{1cm}\textbf{(2.3):}\hspace{0.5mm}$The type (1) and type (2) conditional messages samples by equation(A.3),
(A.4),(A.7),(A.8).And the type (3), type (4) conditional messages samples by equation(A.5),(A.6),(A.9),(A.10).For example we randomly generate a random number $\mathcal{R}$. If$\hspace{1mm}\mathcal{R}<p_{i\rightarrow k\backslash j}^{(0,0)}\hspace{1mm}$(it is calculated by the equation(A.7)), then we select $p_{k\rightarrow i}( c_{k}, c_{i}|0,0)$ as the insert message, otherwise we select $p_{k\rightarrow i}( c_{k}, c_{i}|1,0)$ as the insert message. And we randomly generate a random number $\mathcal{R}$ again. if$\hspace{1mm}\mathcal{R}<p_{i\rightarrow k\backslash j}^{(1,0)}\hspace{1mm}$(it is calculated by the equation(A.8)), then we select $p_{k\rightarrow i}( c_{k}, c_{i}|0,1)$ as the insert message, otherwise we select $p_{k\rightarrow i}( c_{k}, c_{i}|1,1)$ as the insert message.\\

$\textbf{(3)\hspace{0.5mm}Calculate thermodynamic quantity:\hspace{0.5mm}}$ After updating we randomly select k average messages from the population. $\textcircled{1}$we randomly generate a random number $\mathcal{R}$, If$\hspace{1mm}\mathcal{R}<p_{i\rightarrow j}^{(0,0)}\hspace{1mm}$(it is calculated by the equation(A.3)), then we select $p_{j\rightarrow i}( c_{j}, c_{i}|0,0)$ as the insert message, otherwise we select $p_{j\rightarrow i}( c_{j}, c_{i}|1,0)$ as the insert message.\\
$\textcircled{2}$we randomly generate a random number $\mathcal{R}$ again, If$\hspace{1mm}\mathcal{R}<p_{i\rightarrow j}^{(1,0)}\hspace{1mm}$(it is calculated by the equation(A.5)), then we select $p_{j\rightarrow i}( c_{j}, c_{i}|0,1)$ as the insert message, otherwise we select $p_{j\rightarrow i}( c_{j}, c_{i}|1,1)$ as the insert message.\\
$\textcircled{3}$ After this we can determine the another variable$(k,i)$'s state by equation(A.7-A.10) to get the insert messages.In the same way we can determine the other variable's state and insert messages. we can calculate the mean free energy of node $i(<f_{i}>)$ by equation (22) and the general free energy$(g_{i})$ by equation (27).\\
$\textcircled{4}$Randomly choose two messages and calculate the general free energy$g_{(i,j)}$ of edge$(i,j)$ by the equation(26).and select two conditional messages by equation(24) to calculate the mean free energy$<f_{(i,j)}>$ of edge $(i,j)$.\\
$\textcircled{5}$Finally we can calculate the mean free energy density, grand free energy density and complexity by the equations$(28,29,30)$.\\
In the simulation,we update the population $M_{I}=100000$ times to reach the stable point of the population,and to sample $M_{S}=100000$ times to get the condensation transition points of the regular random graph, we can reach the stable point of population by using small update numbers when variable degree is big.But we update the population $M_{I}=5000$ times to reach the stable point of the population,and to sample $M_{S}=5000$ times to get the cluster transition points of the regular random graph, so the cluster transition point is only correct in the range of $\bigtriangledown y=0.05$. The population size is $N=10000$.If we increase the population size, it no make sense to the simulation results.But if we increase the samples number $M_{S}$ of the thermodynamic quantities,the result better than before.In the range $\bigtriangledown y=0.05$,we also can get good results with small update times such as several thousands time.But in the range $\bigtriangledown y=0.01$, we need more and more update times to get good results.
\section{Zero temperature mean field theory }
In this section,Firstly we introduce the Belief Propagation at $\beta=\infty$ which called Warning Propagation. Even the Warning Propagation converge very fast, but it only converge in $C<2.41$ on ER random network, so we must further consider the one step replica symmetry breaking case at $\beta=\infty$. Secondly we derive the Survey Propagation(SP) for the $\beta=\infty$ case, and estimating the ground state energy. At last we predict the energy density by Survey Propagation Decimation(SPD) method.We find that the SP results fit in with SPD and SPD results good as BPD results.
\subsection{Warning Propagation }
In order to estimate the minimal energy of the MDS we must consider the limit property at $\beta=\infty$. There are three cases for the single node(1)the node $i$ appear in the all MDS or $p_{i}=1$.(2)the node $i$ not appear in the all MDS or $p_{i}=0$.(3)the node $i$ appear in some MDS or $p_{i}=0.5$. So There are nine cases for the pair node (i,j),But only the four cavity messages for these cases are possible (1)the node $i$ in the pair nodes $(i,j)$ appears in the all MDS or $(p_{i\rightarrow j}^{1,0}=p_{i\rightarrow j}^{1,1}=0.5;p_{i\rightarrow j}^{0,0}=p_{i\rightarrow j}^{0,1}=0.0)$.(2)the node $i$ in the pair nodes $(i,j)$ not appear in the all MDS and the node $j$ appear in some MDS or $(p_{i\rightarrow j}^{1,0}=p_{i\rightarrow j}^{1,1}=0.0;p_{i\rightarrow j}^{0,0}=p_{i\rightarrow j}^{0,1}=0.5)$.(3)the node $i$ in the pair nodes $(i,j)$ not appear in the all MDS and the node $j$ appear in all MDS or $(p_{i\rightarrow j}^{1,0}=p_{(i,j)}^{1,1}=0.0;p_{i\rightarrow j}^{0,0}=0,p_{i\rightarrow j}^{0,1}=1.0)$.(4)the node $i$ in the pair nodes $(i,j)$ appear in some MDS and the node $j$ appear in some MDS too or $(p_{i\rightarrow j}^{1,0}=p_{i\rightarrow j}^{1,1}=p_{(i,j)}^{0,0}=p_{i\rightarrow j}^{0,1}=0.25)$.\\
There is one warning message $p_{i\rightarrow j}^{0}=1$ for single node, But there are two warning messages for pair nodes$(i,j)$ as $(p_{i\rightarrow j}^{0,1}=1.0,p_{i\rightarrow j}^{0,1}=0.5)$

\begin{tequation}p_{i\rightarrow j}^{0,1}=
\begin{cases}
0.0&\qquad\sum\limits_{k\in\partial i\backslash j}\delta_{p_{k\rightarrow i}^{0,1}}^{1}\geq2\\
0.25&\qquad\sum\limits_{k\in\partial i\backslash j}\delta_{p_{k\rightarrow i}^{0,1}}^{1}=1\\
0.5&\qquad\sum\limits_{k\in\partial i\backslash j}\delta_{p_{k\rightarrow i}^{0,1}}^{1}=0\qquad and \qquad\sum\limits_{k\in\partial i\backslash j}\delta_{p_{k\rightarrow i}^{0,1}}^{0.5}<k-1\\
1.0&\qquad\sum\limits_{k\in\partial i\backslash j}\delta_{p_{k\rightarrow i}^{0,1}}^{1}=0\qquad and \qquad\sum\limits_{k\in\partial i\backslash j}\delta_{p_{k\rightarrow i}^{0,1}}^{0.5}=k-1
\end{cases}
\end{tequation}

equation(31) is called warning propagation equation.If we find the stable point of the warning propagation,then we can calculate the coarse-grained
state of every node as
\begin{equation}p_{i}^{1}=
\begin{cases}
0.0&\qquad\sum\limits_{j\in\partial i}\delta_{p_{j\rightarrow i}^{0,1}}^{1}=0\\
0.5&\qquad\sum\limits_{j\in\partial i}\delta_{p_{j\rightarrow i}^{0,1}}^{1}=1\\
1.0&\qquad\sum\limits_{j\in\partial i}\delta_{p_{j\rightarrow i}^{0,1}}^{1}\geq2
\end{cases}
\end{equation}
and the free energy in general case
\begin{equation}
\begin{split}
F_{i}&=-\frac{1}{\beta}\ln\{\prod_{j\in\partial i}(p_{j\rightarrow i}^{(0,0)}+p_{j\rightarrow i}^{(1,0)})\\
&+e^{-\beta}\prod_{j\in\partial i}(p_{j\rightarrow i}^{(0,1)}+p_{j\rightarrow i}^{(1,1)})-\prod_{j\in\partial i}p_{j\rightarrow i}^{(0,0)}\}
\end{split}
\end{equation}

\begin{equation}
\begin{split}
F_{ij}&=-\frac{1}{\beta}\ln(p_{i\rightarrow j}^{(0,0)}p_{j\rightarrow i}^{(0,0)}+p_{i\rightarrow j}^{(0,1)}p_{j\rightarrow i}^{(1,0)}\\
&\quad+p_{i\rightarrow j}^{(1,0)}p_{j\rightarrow i}^{(0,1)}+p_{i\rightarrow j}^{(1,1)}p_{j\rightarrow i}^{(1,1)})
\end{split}
\end{equation}
from these above two questions we can write the free energy of $\beta=\infty$ as
\begin{tequation}
\begin{split}
E_{min}&=\lim\limits_{\beta\rightarrow\infty}F_{0}=\sum\limits_{i=1}^{N}[\Theta(\sum\limits_{j\in\partial i}\delta_{p_{j\rightarrow i}^{0,1}}^{1}-1)+\delta(\sum\limits_{j\in\partial i}\delta_{p_{j\rightarrow i}^{0,1}}^{0.5},k)]\\
&-\sum\limits_{(i,j)\in w}[(\delta_{p_{i\rightarrow j}^{0,1}}^{1}+\delta_{p_{i\rightarrow j}^{0,1}}^{0.5})*(\delta_{p_{j\rightarrow i}^{0,1}}^{1}+\delta_{p_{j\rightarrow i}^{0,1}}^{0.5})-\delta_{p_{i\rightarrow j}^{0,1}}^{0.5}\delta_{p_{j\rightarrow i}^{0,1}}^{0.5}]
\end{split}
\end{tequation}
The warning propagation gives same results with Replica Symmetry theory,but it can't converge when mean variable degree bigger than 2.41 on ER random graph.
\subsection{Coarse-Grain Survey Propagation }
In order to get the survey propagation we must to know the form of free energy$F_{i\rightarrow j}$, at zero
temperature, from the general form we can derive the free energy$F_{i\rightarrow j}$ as
\begin{equation}
\begin{split}
F_{i\rightarrow j}=&-\frac{1}{\beta}\ln\{2\prod_{k\in\partial i\backslash j}(p_{k\rightarrow i}^{(0,0)}+p_{k\rightarrow i}^{(1,0)})\\
&+2e^{-\beta}\prod_{k\in\partial i\backslash j}(p_{k\rightarrow i}^{(0,1)}+p_{k\rightarrow i}^{(1,1)})-\prod_{k\in\partial i\backslash j}p_{k\rightarrow i}^{(0,0)}\}
\end{split}
\end{equation}
\begin{equation}
F_{i\rightarrow j}=\Theta(\sum\limits_{k\in\partial i\backslash j}\delta_{p_{k\rightarrow i}^{0,1}}^{1}-1)
\end{equation}
the survey propagation for general case as
\begin{tequation}
P_{i\rightarrow j}(p)=\frac{\prod_{k\in\partial i\backslash j}\int\mathcal{D}p_{k\rightarrow i}P_{k\rightarrow i}(p)e^{-yf_{i\rightarrow j}}\delta(p_{i\rightarrow j}-I_{i\rightarrow j}[p_{\partial i\backslash j}])}{\prod_{k\in\partial i\backslash j}\int\mathcal{D}p_{k\rightarrow i}P_{k\rightarrow i}(p)e^{-yf_{i\rightarrow j}}}
\end{tequation}
We can get the Survey Propagation at zero temperature by using the upper two equations (37,38) as
\begin{tequation}
P_{i\rightarrow j}(\delta_{p_{j\rightarrow i}^{0,1}}^{1})=\frac{\prod\limits\limits_{k\in\partial i\backslash j}P_{k\rightarrow i}(\delta_{p_{k\rightarrow i}^{0,1}}^{0.5})}{\prod\limits\limits_{k\in\partial i\backslash j}[1-P_{k\rightarrow i}(\delta_{p_{k\rightarrow i}^{0,1}}^{1})]+e^{-y}P_{01}^{1}}
\end{tequation}
where$P_{01}^{1}=1-\prod\limits\limits_{k\in\partial i\backslash j}[1-P_{k\rightarrow i}(\delta_{p_{k\rightarrow i}^{0,1}}^{1})]$
\begin{tequation}
P_{i\rightarrow j}(\delta_{p_{j\rightarrow i}^{0,1}}^{0.5})=\frac{\prod\limits\limits_{k\in\partial i\backslash j}[1-P_{k\rightarrow i}(\delta_{p_{k\rightarrow i}^{0,1}}^{1})]-\prod\limits\limits_{k\in\partial i\backslash j}P_{k\rightarrow i}(\delta_{p_{k\rightarrow i}^{0,1}}^{0.5})}{\prod\limits\limits_{k\in\partial i\backslash j}[1-P_{k\rightarrow i}(\delta_{p_{k\rightarrow i}^{0,1}}^{1})]+e^{-y}P_{01}^{1}}
\end{tequation}
these two equations are the survey propagation of zero temperature.In the same way we can derive the free energy of node $i$ and edge $(i,j)$ as
\begin{equation}
f_{i}=\Theta(\sum\limits_{j\in\partial i}\delta_{p_{j\rightarrow i}^{0,1}}^{1}-1)+\delta(\sum\limits_{j\in\partial i}\delta_{p_{j\rightarrow i}^{0,1}}^{0.5},k)
\end{equation}
\begin{tequation}
f_{(i,j)}=\sum\limits_{(i,j)\in w}[(\delta_{p_{i\rightarrow j}^{0,1}}^{1}+\delta_{p_{i\rightarrow j}^{0,1}}^{0.5})*(\delta_{p_{j\rightarrow i}^{0,1}}^{1}+\delta_{p_{j\rightarrow i}^{0,1}}^{0.5})-\delta_{p_{i\rightarrow j}^{0,1}}^{0.5}\delta_{p_{j\rightarrow i}^{0,1}}^{0.5}]
\end{tequation}
and the grand free energy of node $i$ and edge $(i,j)$ for the general case as
\begin{equation}
g_{i}=\frac{1}{y}\ln[\prod_{j\in\partial i}\int\mathcal{D}p_{i\rightarrow j}P_{i\rightarrow j}(p)e^{-yf_{i}}]
\end{equation}
\begin{tequation}
g_{(i,j)}=\frac{1}{y}\ln[\int\int\mathcal{D}p_{i\rightarrow j}\mathcal{D}p_{j\rightarrow i}P_{i\rightarrow j}(p)P_{j\rightarrow i}(p)e^{-yf_{(i,j)}}]
\end{tequation}
from the upper four equations we can derive the grand free energy of node $i$ and edge $(i,j)$ at zero temperature as
\begin{equation}
\begin{split}
G_{i}&=-\frac{1}{y}\sum\limits_{i=1}^{N}\ln\{(1-e^{-y})\prod_{j\in\partial i}[1-P_{j\rightarrow i}(\delta_{p_{j\rightarrow i}^{0,1}}^{1})]\\
&+e^{-y}-(1-e^{-y})\prod_{j\in\partial i}P_{j\rightarrow i}(\delta_{p_{j\rightarrow i}^{0,1}}^{0.5})\}
\end{split}
\end{equation}
\begin{equation}
G_{(i,j)}=-\frac{1}{y}\sum\limits_{i=1}^{N}\ln\{1-(1-e^{-y})p_{i\rightarrow j}^{0,1}[nh]
\end{equation}
where
\begin{equation}
\begin{split}
p_{i\rightarrow j}^{0,1}[nh]=&[P_{i\rightarrow j}(\delta_{p_{i\rightarrow j}^{0,1}}^{0.5})+P_{i\rightarrow j}(\delta_{p_{i\rightarrow j}^{0,1}}^{1})]\\
&\times[P_{j\rightarrow i}(\delta_{p_{j\rightarrow i}^{0,1}}^{0.5})+P_{j\rightarrow i}(\delta_{p_{j\rightarrow i}^{0,1}}^{1})]\\
&-P_{i\rightarrow j}(\delta_{p_{i\rightarrow j}^{0,1}}^{0.5}) P_{j\rightarrow i}(\delta_{p_{j\rightarrow i}^{0,1}}^{0.5})
\end{split}
\end{equation}
the grand free energy density as
\begin{equation}
g_{0}\equiv \frac{G_{0}}{N}=\frac{\sum_{i}g_{i}-\sum_{(i,j)}g_{(i,j)}}{N}
\end{equation}
The free energy of the macro state $\alpha$ when $\beta=\infty$ equals to the ground state energy $E_{min}$. The macroscopic average minimal energy $<E_{\beta=\infty}>$ is calculated by the following equation
\begin{tequation}
\begin{split}
&<E_{\beta=\infty}>=\frac{\partial (yG_{0})}{\partial y}\\
&=\sum\limits_{i=1}^{N}\frac{e^{-y}\{1-\prod\limits\limits_{j\in\partial i}[1-P_{j\rightarrow i}(\delta_{p_{j\rightarrow i}^{0,1}}^{1})]+\prod\limits\limits_{j\in\partial i}P_{j\rightarrow i}(\delta_{p_{j\rightarrow i}^{0,1}}^{0.5})\}}{e^{-y}+(1-e^{-y})\{\prod\limits\limits_{j\in\partial i}[1-P_{j\rightarrow i}(\delta_{p_{j\rightarrow i}^{0,1}}^{1})]-\prod\limits\limits_{j\in\partial i}P_{j\rightarrow i}(\delta_{p_{j\rightarrow i}^{0,1}}^{0.5})\}}\\
&-\sum\limits_{(i,j)\in\partial w}^{N}\frac{e^{-y}p_{i\rightarrow j}^{0,1}[nh]}{1-(1-e^{-y})p_{i\rightarrow j}^{0,1}[nh]}
\end{split}
\end{tequation}
we can study the ensemble average property of the MDS problem by using population dynamics with equations(39-42,45,46).In the figure 5 we show the ensemble average one step replica symmetry breaking population dynamics results for MDS problem on ER random graph which mean connectivity C=5,the complexity $\sum=0$ at $y=0$,and then the complexity is not monotonic function of parisi parameter $y$,it increase with the increase of parisi parameter $y$ and reach the maximum value when $y\approx 3.5$.Then the complexity begin to decline with the increase of $y$ and change to negative when $y\approx 7.3$.From the figure 5 we can see that there are two parts of the complexity graph when it is a function of energy,but because of only the concave part is decline function of energy, so it has the physical meaning.And the grand free energy not monotonic function of $y$,it reach the biggest point when the complexity change to negative at $y\approx 7.3$.So the corresponding minimal energy density $u=0.2068$ is the minimum energy density (ground state energy) of MDS problem at this mean connectivity.\\
\begin{figure}[htb]
  \centering
  \includegraphics[width=8cm,height=5cm]{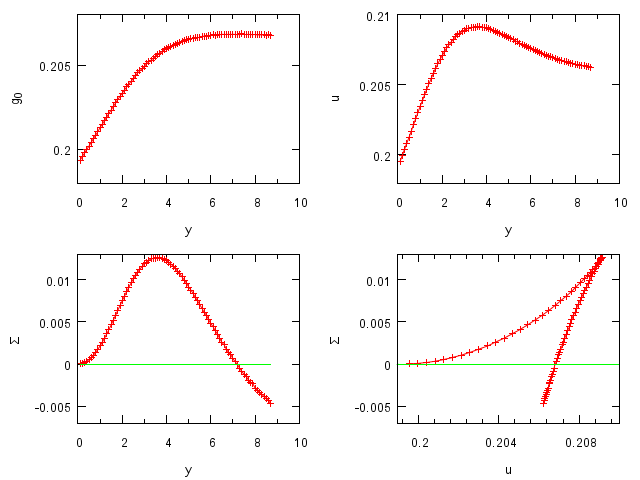}
  \caption{Survey Propagation results of zero temperature MDS problem on ER random graph and with mean connectivity c=5 by population dynamics,
  in the first three graph,x-axis denotes the parisi parameter Y, y-axis denotes the thermodynamic quantities.The complexity equals to zero when parisi parameter y roughly equal to 7.23,at this point,we select the corresponding energy as ground state energy which equal to 0.2068.In the right down graph,the x-axis denotes the energy density and y-axis denotes the complexity.}
\end{figure}
We can calculate the some microscopic statistical quantities by using zero temperature equations (39,40),for example the probability(statistical total weight of all macro state) of the variable stay in coarse grained state,we use $p_{i}(0)$ to denotes the probability of the variable stay in totally not be covered state,$p_{i}(1)$denotes the probability of the variable stay in totally be covered state,and $p_{i}(*)$ denotes the probability of the variable stay in not be freezing(in some micro state be covered) state,we can derive the representation of these three probabilities by one step replica symmetry breaking mean field theory as
\begin{tequation}
\begin{split}
P_{i}(0)=&\frac{\prod_{j\in\partial i}[1-P_{j\rightarrow i}(\delta_{p_{j\rightarrow i}^{0,1}}^{1})]-\prod_{j\in\partial i}P_{j\rightarrow i}(\delta_{p_{j\rightarrow i}^{0,1}}^{0.5})}{\prod_{j\in\partial i}[1-P_{j\rightarrow i}(\delta_{p_{j\rightarrow i}^{0,1}}^{1})]-\prod_{j\in\partial i}P_{j\rightarrow i}(\delta_{p_{j\rightarrow i}^{0,1}}^{0.5})}\\
&\frac{}{+e^{-y}\{1-\prod_{j\in\partial i}[1-P_{j\rightarrow i}(\delta_{p_{j\rightarrow i}^{0,1}}^{1})]+\prod_{j\in\partial i}P_{j\rightarrow i}(\delta_{p_{j\rightarrow i}^{0,1}}^{0.5})\}}
\end{split}
\end{tequation}
\begin{tequation}
\begin{split}
P_{i}(*)=&\frac{\sum\limits_{j\in\partial i}P_{j\rightarrow i}(\delta_{p_{j\rightarrow i}^{0,1}}^{1})\prod_{k\in\partial i\backslash j}[1-P_{k\rightarrow i}(\delta_{p_{k\rightarrow i}^{0,1}}^{1})]}{\prod_{j\in\partial i}[1-P_{j\rightarrow i}(\delta_{p_{j\rightarrow i}^{0,1}}^{1})]-\prod_{j\in\partial i}P_{j\rightarrow i}(\delta_{p_{j\rightarrow i}^{0,1}}^{0.5})}\\
&\frac{}{+e^{-y}\{1-\prod_{j\in\partial i}[1-P_{j\rightarrow i}(\delta_{p_{j\rightarrow i}^{0,1}}^{1})]+\prod_{j\in\partial i}P_{j\rightarrow i}(\delta_{p_{j\rightarrow i}^{0,1}}^{0.5})\}}
\end{split}
\end{tequation}
\begin{equation}
P_{i}(1)=1-P_{i}(0)-P_{i}(*)
\end{equation}
We proceed the one step replica symmetry breaking population dynamics on different mean connectivity ER random graph, and we get the minimal energy density of ER random network ensemble on different mean connectivity $C$ .In the table 2 we list the theoretical computation results of $C\le 10$,we
can see that the transition point $y$ not depend on the mean connectivity $C$.
\begin{table}[!hbp]
\tiny
\caption{the cluster transition point inverse temperature $\beta_{d}$ of ER Random graph}
\begin{tabular}{p{0.45cm}p{0.45cm}p{0.45cm}p{0.45cm}p{0.45cm}p{0.45cm}p{0.45cm}p{0.45cm}p{0.45cm}}
\hline
K & 3 & 4 & 5 & 6&7&8&9&10 \\
\hline
$y^{*}\approx$ & 8.0 & 7.29 & 7.15 & 7.19&7.31&7.47&7.64&7.82 \\
\hline
$u_{min}$ & 0.3176 & 0.2498 & 0.2068 & 0.178&0.1576&0.142&0.130&0.120 \\
\hline
\end{tabular}
\end{table}
\\
In the simulation, we update the population $M_{I}=5000$ times to reach the stable point of the population,and to sample $M_{S}=5000$ times to get the $\sum=0$ points and the corresponding ground energy value $E_{min}$ on the ER random graph, the cluster transition point is only correct in the range of $\bigtriangledown y=0.05$, but the ground energy is correct in the range of $\bigtriangledown E=0.0001$. The population size is $N=100000$.If we increase the updating number and sampling number, it no make sense to the simulation results.But if we increase the population size $N$ of the thermodynamic quantities,the result better than before.In the range $\bigtriangledown y=0.05$,we also can get good results with small update times such as several thousands time.But in the range $\bigtriangledown y=0.01$, we need more and more update times to get good results.Our updating and sampling number increase with the decreasing of variable degree.
\subsection{Survey Propagation Decimation}
we still can study the statistical property of microscopic configuration on single network system by survey propagations(39,40),and the survey propagation easy to find the stable point of a given network when the parisi parameter $y$ small enough, and then we can calculate the thermodynamic quantities by equations(41,42,45-47), but the survey propagation not converge any more when $y$ bigger enough.For example our simulation results indicate that the survey propagation not converge when $y\ge 2.1$ on $C=10$ ER network.The reason of not convergence is that the coarse grained assumption not very good any more on the microscopic configuration space that energy close to ground state energy,it is need to use more detailed coarse grained assumption.The other more intrinsical reason is that the one step replica symmetry breaking mean field theory is not good enough to describe the microscopic configuration space that energy close to ground state energy,it is need to consider more high step expansion of partition function.There are more articles about convergence of coarse grained survey propagation\cite{22,23}.\\
It is possible to construct one or more close to optimal MDS solution for a given graph $W$ by one step replica symmetry breaking mean field theory.One of the very efficient algorithm is the Survey Propagation-guided Decimation algorithm\cite{24}.The core idea of this algorithm is to determine the probability of be freezing by equation(50-52), and select the small part of variables that has biggest be covering probability to set the be covering probability $ c_{i}=1$,so then simplify the network step by step.Now we introduce the concrete procedure of this algorithm in detail:\\
(1) Read in the network $W$, setting the covering probability of every vertex as uncertain,and to define four coarse grained messages $P_{i\rightarrow j}(\delta_{p_{i\rightarrow j}^{0,1}}^{0.5}),P_{i\rightarrow j}(\delta_{p_{i\rightarrow j}^{0,1}}^{1})$,and $P_{j\rightarrow i}(\delta_{p_{j\rightarrow i}^{0,1}}^{0.5}),P_{j\rightarrow i}(\delta_{p_{j\rightarrow i}^{0,1}}^{1})$ on every edge of the given graph.Randomly initialize the every message in the range(0,1],but the every group two messages $P_{i\rightarrow j}(\delta_{p_{i\rightarrow j}^{0,1}}^{0.5})$
,$P_{i\rightarrow j}(\delta_{p_{i\rightarrow j}^{0,1}}^{1})$ must satisfy the normalization condition$P_{i\rightarrow j}(\delta_{p_{i\rightarrow j}^{0,1}}^{1})$+$P_{i\rightarrow j}(\delta_{p_{i\rightarrow j}^{0,1}}^{0.5})$
$+P_{i\rightarrow j}(\delta_{p_{i\rightarrow j}^{0,1}}^{0.25})$+$P_{i\rightarrow j}(\delta_{p_{i\rightarrow j}^{0,1}}^{0})=1$. Appropriately setting the macroscopic inverse temperature $y$,it is good to set closet value of biggest convergence value of macroscopic inverse temperature,for example we select $y=3$ if the survey propagation can not converge when$y\ge 3.01$.\\
(2)Iterate $L_{0}$ steps (for example $L_{0}=100$) the coarse grained survey propagation equations(39,40) try to converge to one stable point.Every step of iteration, to select one node $i$ and updating every corresponding messages of node $i$.After updating every node messages $L_{0}$ times we can calculate the coarse grained probability $(P_{i}(1),P_{i}(*),P_{i}(0))$ by equations $(50-52)$.\\
(3)Ordering all the variables that not be freezing with the value of $(P_{i}(1)$ from high to low.Select the foremost $r$ (for example $r=0.01$) percent to set the covering state $ c_{i}=1$ and adding these variables to MDS.\\
(4)Then simplify network by deleting all the edges between observed nodes and deleting all the be occupied variables.If the rest network still contain one or more leaf node\cite{2} or connectivity $d_{i}=1$ nodes, then we procedure GLR process\cite{1} until no exist leaf nodes in the rest network,and then simplify the network again.Iterating this procedure (simplify-GLR-simplify) until the rest network not contains any leaf nodes.If the rest network not contains any nodes and edges,then stop the program and output the MDS.\\
(5)If the rest network still contains some nodes,then Iterating the survey propagation(39,40) $L_{1}$ (for example $L_{1}=10$) steps,and then repeat the step (3),(4),(5).\\
The figure 6 shows the numerical results of survey propagation decimation algorithm on ER random graph. We can see that the SPD results very close to BPD results,it is explain that the SPD algorithm can find the very closet optimal solution.We perform the BPD algorithm by the way mentioned in the reference\cite{1}.

\begin{figure}[htbp]
  \centering
  \includegraphics[width=8cm,height=5cm]{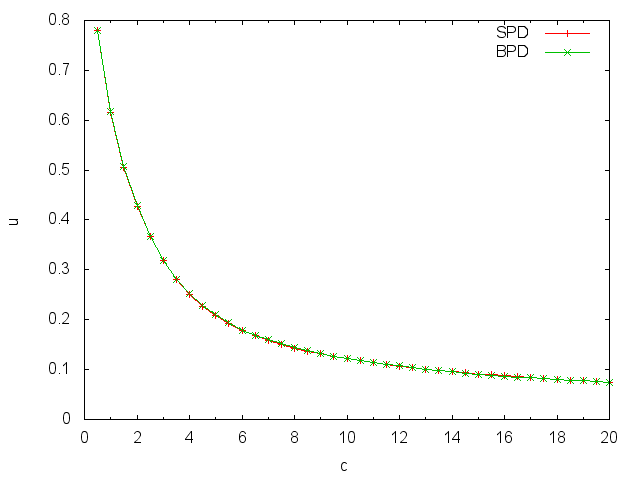}
  \caption{The solid line is the result of the SPD, and the cross point line is the result of BPD. Our simulation performs on the ER random graph (include $10^4$ variables).}
\end{figure}
\section{$\hspace{1.5mm}$Discussion}
In this work, firstly we derive the one step replica symmetry breaking equations of MDS problem at $y=\beta$, we find the condensation transition point and cluster transition point of regular random graph. The corresponding energy of the cluster transition point inverse temperature $\beta_{d}$ equals to the threshold value $x_{c}$, namely $E_{\beta_{d}}=x_{c}$. The complexity graph of MDS problem likes to three body interaction spin glass model and 4-sat problem\cite{20,21}, but the change rules of these two transition points with the variable degree is totally different with the three body interaction spin glass model. In the three body interaction spin glass model, these two transition points always decline with the increasing of variable degree, but these are not monotonic function of variable degree in MDS problem.Secondly we derive the warning propagation and prove that the warning propagation equation only converge when the network no contain any core\cite{1}.There are only one warning on vertex cover problem\cite{3},but the MDS problem have two warnings.Then we derive the Survey Propagation function of zero temperature to find the ground state energy and the corresponding transition point of macroscopic inverse temperature,the change rules of the transition point like with vertex cover problem. The corresponding energy of the transition point parisi parameter $Y$ equals to the threshold value $x_{c}$, namely $E_{Y}=x_{c}$. And then we implement the survey propagation decimation (SPD) algorithm at zero temperature to estimate the size of MDS, it's result same good as BPD results. \\
Before we have studied the MDS problem on undirected network and directed network by statistical physics,and now we study the undirected MDS problem by one step replica symmetry breaking mean field theory.Later we continue to study the rest events such as directed MDS problem under one step replica symmetry breaking mean field case and long range frustration theory on the MDS problem and so on.
\section{$\hspace{2mm}$ Acknowledgement}
Yusupjan.Habibulla very thanks Prof Haijun.Zhou for helpful discussion, guidance and support. Yusupjan.Habibulla still thanks Pan.Zhang, Shaoming.Qin for very helpful discussion, we procedure numerical simulation on the cluster of Prof Haijun.Zhou's group, research partially supported by the Doctor startup fund of Xinjiang University of China (grant number 208-61357)

\begin{appendix}

\setcounter{equation}{0}
\renewcommand{\theequation}{A.\arabic{equation}}
\section{samples by probability function (18)}
We use equation (18) to discuss the sample scheme. Our problem is to give a spin value to the around variables of the node $i$,and the joint probability of these spin values obey the equation (18).The directly way of sampling is to calculate $2^{k}$ different kind of spin value configurations probability by the equation (18),then generate one random number in the range (0,1),and then corresponding to the random number to select one spin value configuration.In the directly sampling way,we need to calculate the probability of all the configurations,so this way not good when $k\gg 1$.\\
The another convenient way of sampling is sequential sampling,in this way we firstly determine the value $c_{i}$ of the node $i$,then corresponding to this value $c_{i}$ to determine the value $c_{j}$ of node $j$, and then we correspond to these value $c_{i},c_{j}$ to calculate the value $c_{k}$ of node $k$, and so on until determine all the value of the around nodes of node $i$.\\
Firstly,we use equation (18) to derive the marginal probability of the node $i$ as
\begin{tequation}
p_{i}^{(0)}=\frac{\prod_{j\in i}(\bar{p}_{j\rightarrow i}^{(1,0)}+\bar{p}_{j\rightarrow i}^{(0,0)})-\prod_{j\in i}\bar{p}_{j\rightarrow i}^{(0,0)}}{\prod_{j\in i}(\bar{p}_{j\rightarrow i}^{(1,0)}+\bar{p}_{j\rightarrow i}^{(0,0)})+e^{-\beta}\prod_{j\in i}(\bar{p}_{j\rightarrow i}^{(1,1)}+\bar{p}_{j\rightarrow i}^{(0,1)})-\prod_{j\in i}\bar{p}_{j\rightarrow i}^{(0,0)}}
\end{tequation}
\begin{tequation}
p_{i}^{(1)}=\frac{e^{-\beta}\prod_{j\in i}(\bar{p}_{j\rightarrow i}^{(1,1)}+\bar{p}_{j\rightarrow i}^{(0,1)})}{\prod_{j\in i}(\bar{p}_{j\rightarrow i}^{(1,0)}+\bar{p}_{j\rightarrow i}^{(0,0)})+e^{-\beta}\prod_{j\in i}(\bar{p}_{j\rightarrow i}^{(1,1)}+\bar{p}_{j\rightarrow i}^{(0,1)})-\prod_{j\in i}\bar{p}_{j\rightarrow i}^{(0,0)}}
\end{tequation}
corresponding to the above equation we can generate the spin value of the node $i$.After given the spin value of the node $i$, we can derive the conditional probability of the node $j$ as
\begin{tequation}
p_{i\rightarrow j}^{(0,0)}=\frac{\bar{p}_{j\rightarrow i}^{(0,0)}\times\prod_{k\in i\backslash j}(\bar{p}_{k\rightarrow i}^{(1,0)}+\bar{p}_{k\rightarrow i}^{(0,0)})-\prod_{k\in i}\bar{p}_{k\rightarrow i}^{(0,0)}}{\prod_{k\in i}(\bar{p}_{k\rightarrow i}^{(1,0)}+\bar{p}_{k\rightarrow i}^{(0,0)})-\prod_{k\in i}\bar{p}_{k\rightarrow i}^{(0,0)}+e^{-\beta}\prod_{k\in i}(\bar{p}_{k\rightarrow i}^{(0,1)}+\bar{p}_{k\rightarrow i}^{(1,1)})}
\end{tequation}
\begin{tequation}
p_{i\rightarrow j}^{(0,1)}=\frac{\bar{p}_{j\rightarrow i}^{(1,0)}\times\prod_{k\in i\backslash j}(\bar{p}_{k\rightarrow i}^{(1,0)}+\bar{p}_{k\rightarrow i}^{(0,0)})}{\prod_{k\in i}(\bar{p}_{k\rightarrow i}^{(1,0)}+\bar{p}_{k\rightarrow i}^{(0,0)})-\prod_{k\in i}\bar{p}_{k\rightarrow i}^{(0,0)}+e^{-\beta}\prod_{k\in i}(\bar{p}_{k\rightarrow i}^{(0,1)}+\bar{p}_{k\rightarrow i}^{(1,1)})}
\end{tequation}
\begin{tequation}
p_{i\rightarrow j}^{(1,0)}=\frac{\bar{p}_{j\rightarrow i}^{(0,1)}\times e^{-\beta}\prod_{k\in i\backslash j}(\bar{p}_{k\rightarrow i}^{(0,1)}+\bar{p}_{k\rightarrow i}^{(1,1)})}{\prod_{k\in i}(\bar{p}_{k\rightarrow i}^{(1,0)}+\bar{p}_{k\rightarrow i}^{(0,0)})-\prod_{k\in i}\bar{p}_{k\rightarrow i}^{(0,0)}+e^{-\beta}\prod_{k\in i}(\bar{p}_{k\rightarrow i}^{(0,1)}+\bar{p}_{k\rightarrow i}^{(1,1)})}
\end{tequation}
\begin{tequation}
p_{i\rightarrow j}^{(1,1)}=\frac{\bar{p}_{j\rightarrow i}^{(1,1)}\times e^{-\beta}\prod_{k\in i\backslash j}(\bar{p}_{k\rightarrow i}^{(0,1)}+\bar{p}_{k\rightarrow i}^{(1,1)})}{\prod_{k\in i}(\bar{p}_{k\rightarrow i}^{(1,0)}+\bar{p}_{k\rightarrow i}^{(0,0)})-\prod_{k\in i}\bar{p}_{k\rightarrow i}^{(0,0)}+e^{-\beta}\prod_{k\in i}(\bar{p}_{k\rightarrow i}^{(0,1)}+\bar{p}_{k\rightarrow i}^{(1,1)})}
\end{tequation}
and then we can calculate the conditional probability $p(k|i,j)$ of node $k$ when the spin value of nodes $(i,j)$ are given.
\begin{tequation}
p_{i\rightarrow k\backslash j}^{(0,0)}=\frac{\bar{p}_{k\rightarrow i}^{(0,0)}\times\prod_{l\in i\backslash (j,k)}(\bar{p}_{l\rightarrow i}^{(1,0)}+\bar{p}_{l\rightarrow i}^{(0,0)})-\delta_{ c_{j}}^{0}\prod_{l\in i\backslash j}\bar{p}_{l\rightarrow i}^{(0,0)}}{\prod_{l\in i\backslash j}(\bar{p}_{l\rightarrow i}^{(1,0)}+\bar{p}_{l\rightarrow i}^{(0,0)})-\delta_{ c_{j}}^{0}\prod_{l\in i\backslash j}\bar{p}_{l\rightarrow i}^{(0,0)}}
\end{tequation}
\begin{equation}
p_{i\rightarrow k\backslash j}^{(0,1)}=\frac{\bar{p}_{k\rightarrow i}^{(1,0)}\times\prod_{l\in i\backslash j}(\bar{p}_{l\rightarrow i}^{(1,0)}+\bar{p}_{l\rightarrow i}^{(0,0)})}{\prod_{l\in i\backslash j}(\bar{p}_{l\rightarrow i}^{(1,0)}+\bar{p}_{l\rightarrow i}^{(0,0)})-\delta_{ c_{j}}^{0}\prod_{l\in i\backslash j}\bar{p}_{l\rightarrow i}^{(0,0)}}
\end{equation}
\begin{equation}
p_{i\rightarrow k\backslash j}^{(1,0)}=\frac{\bar{p}_{k\rightarrow i}^{(0,1)}\times\prod_{l\in i\backslash (j,k)}(\bar{p}_{l\rightarrow i}^{(0,1)}+\bar{p}_{l\rightarrow i}^{(1,1)})}{\prod_{l\in i\backslash j}(\bar{p}_{l\rightarrow i}^{(0,1)}+\bar{p}_{l\rightarrow i}^{(1,1)})}
\end{equation}
\begin{equation}
p_{i\rightarrow k\backslash j}^{(1,1)}=\frac{\bar{p}_{k\rightarrow i}^{(1,1)}\times\prod_{l\in i\backslash (j,k)}(\bar{p}_{l\rightarrow i}^{(0,1)}+\bar{p}_{l\rightarrow i}^{(1,1)})}{\prod_{l\in i\backslash j}(\bar{p}_{l\rightarrow i}^{(0,1)}+\bar{p}_{l\rightarrow i}^{(1,1)})}
\end{equation}
corresponding the above equations we can generate the spin value of node $k$. And so on,we can generate one configuration of the all spin values.
\end{appendix}

\end{document}